# Random matrix theory analysis of a temperature-related transformation in statistics of Fano-Feshbach resonances in Thulium atoms

E.T. Davletov[1,2], V.V. Tsyganok[1,2,3], V. A. Khlebnikov[1], D. A. Pershin[1,4], A.V Akimov[1,4,5]

[1]Russian Quantum Center, Business Center "Ural", 100A Novaya Street Skolkovo, Moscow, 143025, Russia

[2]Moscow Institute of Physics and Technology, Institutskii pereulok 9, Dolgoprudny, Moscow Region 141701, Russia

[3]National University of Science and Technology MISIS, Leninsky Prospekt 4, Moscow, 119049, Russia,

[4]PN Lebedev Institute RAS, Leninsky Prospekt 53, Moscow, 119991, Russia

[5]Texas A&M University, TAMU 4242, College Station, TX 77843, USA

## I.     ABSTRACT

Recently, transformation from random to chaotic behavior in the statistics of Fano-Feshbach resonances was observed in thulium atoms with rising ensemble temperature. We performed random matrix theory simulations of such spectra and analyzed the resulting statistics. Our simulations show that, when evaluated in terms of the Brody parameter, resonance statistics do not change or change insignificantly with rising temperature if temperature is the only changing parameter. In the experiments evaluated, temperature was changed simultaneously with optical dipole trap depth. Thus, simulations included the Stark shift based on the known polarizability of the free atoms and assuming their polarizability remains the same in the bound state. Somewhat surprisingly, we found that, while including the Stark shift does lead to minor statistical changes, it does not change the resonance statistics and therefore is not responsible for the experimentally observed statistic transformation. This observation suggests that either our assumption regarding the polarizability of Feshbach molecules is poor or that an additional mechanism changes the statistics and leads to more chaotic statistical behavior.

## II.    INTRODUCTION

Fano-Feshbach resonances play an important role in the control of interatomic interactions [1,2]. These resonances enable the scattering lengths of the elastic binary collisions [3–5] to be changed, turning on and off specific interactions, making Fano-Feshbach resonances a key instrument in quantum simulations [6–8]. In the case of lanthanide atoms, Fano-Feshbach resonances have recently attracted a large amount of attention due to theoretical and experimental demonstration of chaotic statistics in the distribution of these resonances [9–11]. Contrary to the case of erbium and dysprosium, in which chaotic behavior is an intrinsic property of the atomic system, thulium atoms demonstrate both random and chaotic statistics, depending on the temperature of the atomic ensemble [10], in experiments performed in an optical dipole trap (ODT) when temperature was changed together with the trap depth.

Changing the temperature results in two effects: 1) on one side, resonance positions shift with temperature [12], and 2) on the other side, new resonances associated with nonzero mutual angular momentum in the open channel appear as the temperature rises [13–15]. Increasing resonance densities could be responsible for the observed change in resonance statistics. Moreover, resonance shifts can change the mutual spacing of the resonances and, therefore, change the statistics. Finally, the fact that the temperature changes simultaneously with ODT depth can cause an additional shift in the resonance position, which can also affect the statistics. Such additional shift in resonance position was indeed detected in previous work [10].

In the present study, we performed random matrix theory (RMT) simulations of the Fano-Feshbach resonances in atomic thulium and investigated the resulting transformations of resonant statistics with changing temperature and ODT power. We show that the temperature-assisted emergence of independent D resonances does not cause random to chaotic statistical changes with rising temperature. Somewhat surprisingly, while demonstrating minor statistical effects within the ODT power range investigated, the resonance shift associated with the Stark effect does not introduce a significant trend and, therefore, cannot be responsible for the experimentally observed transformation toward more chaotic statistics.

## III.  RMT MODEL

The collisions of two atoms were considered in the center of a mass reference frame, where the motion of these two atoms could be presented as a reduced mass that is falling at the origin. This way the collision can naturally be considered from the point of view of the diatomic molecular potential. Due to internal degrees of freedom, namely, components of the total angular momentum $m_F$, the molecule is subject to integration within the external magnetic and electrical fields. In our experiments, each atom was prepared in the lowest energetic state, $m_F = -4$. The Feshbach resonance is observed when the energy of the colliding atoms is equal to the energy of the molecular bound state, corresponding to a different atomic $m_F$ value [16].

Such resonances can be categorized by the relative orbital momentum of the colliding particles, or molecular orbital momentum $l$. Traditionally, for lanthanide atoms [9–11] the resonances are characterized by open channels. Thus, S- resonances correspond to $l = 0$, and D – resonances correspond to next possible state of the bosons partial wave $l = 2$. Due to the so-called "centrifugal barrier" [17], D -resonances appear only at relatively high temperatures, while S- resonances can be observed at effectively low temperatures [10]. With this in mind, we have considered two experimentally observed [10] sets of resonances: 1) the S- resonances obtained at 2 μK and 2) the S+D resonances observed at 12 μK.

RMT is a quite general approach that allows evaluation of the statistical properties of complex system eigenstates [18]. It was initially developed to study complex scattering properties in nuclear physics [19] and was recently used to examine the collisional properties of ultracold lanthanide atoms [9,11,20]. The RMT approach uses an ensemble of random Hamiltonians, instead of an exact Hamiltonian that is unknown, and preserves some of the statistical properties of the energy levels. We use an adaptation of the approach presented in references [9]and [20] to model the molecular bound states and corresponding Feshbach resonances of thulium atoms.

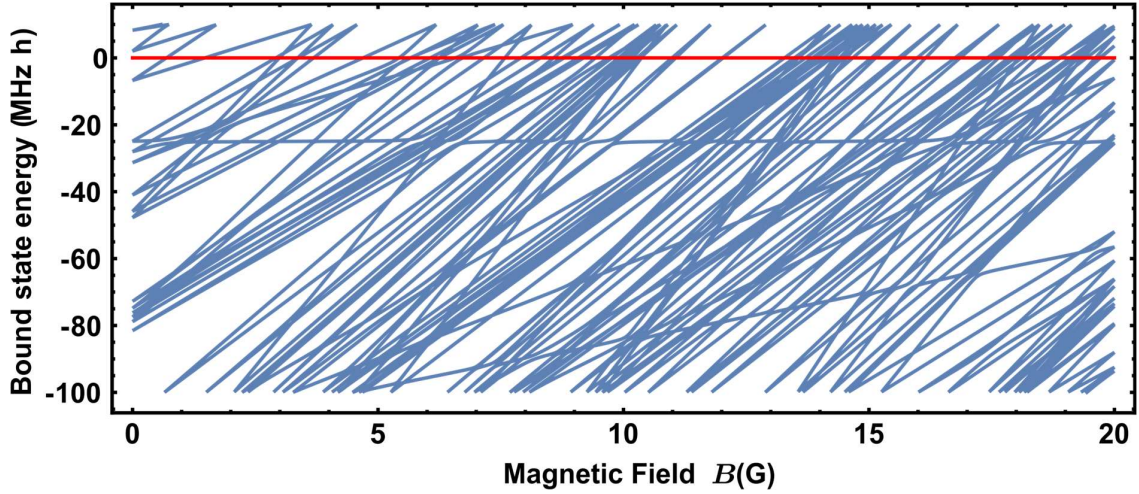

Figure 1 *A characteristic molecular spectrum obtained using the RMT approach. The zero-energy level is taken to be equal to the sum of two atoms in the dipole trap with a projection of magnetic moments* $m_f = -4$.

The molecular Hamiltonian of diatomic Tm[169] molecule is modeled as follows:

$$H = H_b + H_Z + H_S + H_{cpl} \; . \tag{1}$$

Here, $H_b$, $H_z$ and $H_S$ have only diagonal matrix elements. $H_b$ corresponds to the molecular Hamiltonian of two interacting atoms in the free space, while $H_z$ and $H_S$ stand for the Zeeman and Stark shifts, respectively. In conjunction, $H_z$ and $H_S$ shift the specific energy state of the two atoms. $H_{cpl}$ determines the off-diagonal elements and is responsible for anisotropic coupling originating from the dipole-dipole and anisotropic part of the van der Waals dispersion interaction.

Following reference [20], elements of the matrix $H_b$ were randomly sampled so that nearest-neighbor spacings obey Poisson statistics. This assumes no chaos in the initial distribution of energy levels. The mean energy nearest-neighbor spacing of these elements is denoted as $\epsilon_b$. $H_s$ elements are given by:

$$H_s = U(I, m_1^F, F_1) + U(I, m_2^F, F_2) \; , \tag{2}$$

where $F$ is the total atom angular momentum quantum number and $m^F$ is its projection quantum number, $I$ is the peak intensity of the trapping laser beam, and $U$ is the Stark energy shift, which depends on laser beam intensity and geometry as well as atom polarizability [21]. The pairs of $m^F, F$ were sampled from a uniform distribution $[-m^{F=3}, ..., m^{F=3}, -m^{F=4}, ..., m^{F=4}]$ corresponding to all possible total angular momentum projections in both hyperfine sublevels. The Stark energy shift is given by:

$$U(\omega) = -\frac{1}{2\varepsilon_0 c} I(r) \operatorname{Re}[\alpha_{tot}] = U_s + U_\upsilon + U_t$$

$$U_s = -\frac{1}{2\varepsilon_0 c} I(r) \operatorname{Re}[\alpha_s(\omega)]$$

$$U_\upsilon = -\frac{1}{2\varepsilon_0 c} I(r) \varepsilon \cos\theta_k \frac{m_F}{2F} \operatorname{Re}[\alpha_\upsilon(\omega)]$$  (3)

$$U_t = -\frac{1}{2\varepsilon_0 c} I(r) \frac{3m_F^2 - F(F+1)}{F(2F-1)} \cdot \frac{3\cos^2\theta_p - 1}{2} \operatorname{Re}[\alpha_t(\omega)]$$

Here, $\omega$ is trapping light frequency; $\varepsilon_0$ is the vacuum permittivity; $c$ is the speed of light; $I(r)$ is the laser intensity profile; $\varepsilon = |\vec{u}^* \times \vec{u}|$ is the ellipticity parameter with $\vec{u}$ as the normalized Jones vector; $\theta_p = \angle(\vec{E}, \vec{B})$ and $\theta_k = \angle(\vec{k}, \vec{B})$; $\vec{E}$ and $\vec{B}$ are the electrical and magnetic components of the trapping beam light with $\vec{k}$ as its wave vector; $\alpha_{tot}$ is the total atomic polarizability; and $\alpha_s(\omega)$, $\alpha_\upsilon(\omega)$, $\alpha_t(\omega)$ are scalar, vector, and tensor dynamic dipole polarizabilities, respectively. A more detailed description of these quantities and trap geometry can be found in [21].

$H_Z$ elements are given by

$$H_Z = M^F g_F \mu_B B,$$  (4)

where $M^F = m_1^F + m_2^F$, $g_F$ is the Lande g-factor depending on the hyperfine component, $\mu_B$ is the Bohr magneton, and $B$ is the magnetic field value that moves initial and bound states via the Zeeman effect.

$H_{cpl}$ elements were sampled according to normal distribution with zero mean and $v_{cpl}$ variance, where $v_{cpl}$ sets the strength of coupling. With the matrix elements distributed according to the Gaussian orthogonal ensemble, this Hamiltonian aspect introduces chaotic behavior into the distribution of the molecular bound states arising from anisotropic coupling. For each realization of the Hamiltonian $H$, we calculated the corresponding eigenvalues on the magnetic-field from 0 to 20 G, the same region used in experiment [10], and obtain a set of bound-state energies and magnetic field values. The Feshbach resonance position was found by determining the magnetic field at which the energy of the bound state is equal to the entrance channel energy.

The initial state of thulium gas in our trap is $F = 4, m_F = -4$ [22], meaning that the entrance channel energy corresponds to two free $m_F = -4$ atoms. Thus, we follow [9] and subtract the entrance channel energy at each magnetic field value, so that the position of the eigenvalue crossing zero corresponds to the Feshbach resonance position (see Figure 1.).

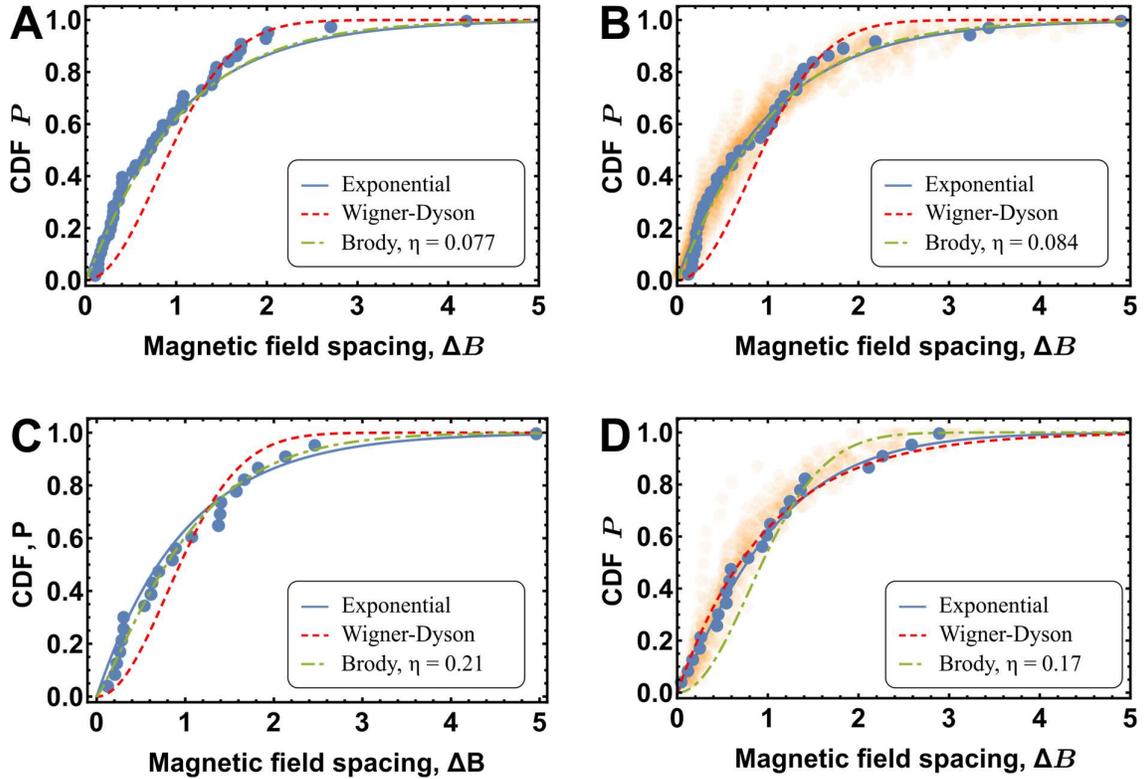

Figure 2. Comparison of NNSFR distributions obtained by RMT simulations for S

and D resonances. (A) ECDF of NNSFR for the spectrum measured at 2.2 μK fitted with Brody distribution ($\eta_S = 0.08$). (B) ECDF of the spectrum generated by RMT simulations with mean energy spacing $\epsilon_S / h = 5.6$ MHz and coupling constant $v_{cpl}^S = 0.7$ MHz were chosen to reproduce the density $\rho_{2.2\mu K}^S$ and Brody distribution constant $\eta_S$ obtained from the experiment. (C) ECDF of NNSFR distribution for the D resonance alone, measured at 12 μK (S resonances subtracted from all observed resonances) and fitted with Brody distribution ($\eta_D = 0.21$). (D) NNSFR ECDF of the spectrum generated by RMT simulations with mean energy spacing $\epsilon_D / h = 10.7$ MHz and coupling constant $v_{cpl}^D / h = 2.9$ MHz, chosen to reproduce the experimental density of D-resonances $\rho_{12\mu K}^D$ and the corresponding Brody parameter $\eta_D$. In B and D the shaded colors around the simulations represent all realizations of RMT simulations, with the number of realizations being proportional to the opacity of the shaded area.

We characterized the chaotic behavior of the system by analyzing nearest-neighbor spacings in the Feshbach resonance (NNSFR) spectra. In previous analyses of distributions of resonance positions and transformation from random to chaotic statistics, the Berry-Robnik distribution [18] was used. However, this distribution employs a complicated analytical expression and, therefore, is not suitable for the analysis below. An alternative approach is to utilize the fully analytical Brody distribution [9,11,20], which can also be used to quantify chaoticity of statistical properties. The Brody distribution is given by [18,23]:

$$P_B(s, \eta) = b(1 + \eta)s^\eta exp(-bs^{\eta+1}) \qquad (5)$$

To gain insight into random to chaos transformations [10] in the statistics of Feshbach spectrum resonances, the S and D resonances were modeled as independent spectra having their own mean energy spacings, $\epsilon_S / h = 5.6$ MHz and $\epsilon_D / h = 10.7$ MHz, respectively, and coupling strength parameters, $v_{cpl}^S = 0.7$ MHz and $v_{cpl}^D / h = 2.9$ MHz. Details of the simulation, such as the number of included levels, discretization, etc., are summarized in Table 1. These parameters were chosen so that the individual spectra generated using the RMT model would reproduce the experimental S and D resonance spectra statistics with regard to nearest neighbor spacings of Feshbach resonances' distribution parameters, such as resonance densities, $\rho_{2.2\mu K}^S = 2$ and $\rho_{12\mu K}^D = 1$, and Brody parameters, $\eta_S = 0.08$ and $\eta_D = 0.21$. The corresponding experimental NNSFR empirical cumulative distribution function (ECDF) obtained for S and D resonances, alone, are shown in Figure 2A and Figure 2C. The corresponding modeled NNSFR distributions are depicted in Figure 2B and Figure 2D and have densities of resonances $\rho_{RMT}^{S} = 2.1 \pm 0.4 G^{-1}$ and $\rho_{RMT}^{D} = 1.0 \pm 0.2 G^{-1}$ and Brody parameters $\eta_{S}^{RMT} = 0.07 \pm 0.03$ and $\eta_{D}^{RMT} = 0.18 \pm 0.05$, respectively.

Mixed spectrum analysis shows that the situation becomes more random, resulting in a Brody parameter of $\eta_{S+D}^{RMT} = 0.07 \pm 0.03$ (see Figure 3B). In contrast, experimental mixed spectra display significantly more chaotic behavior, with a Brody parameter of $\eta_{S+D} = 0.63$ (see Figure 3A).

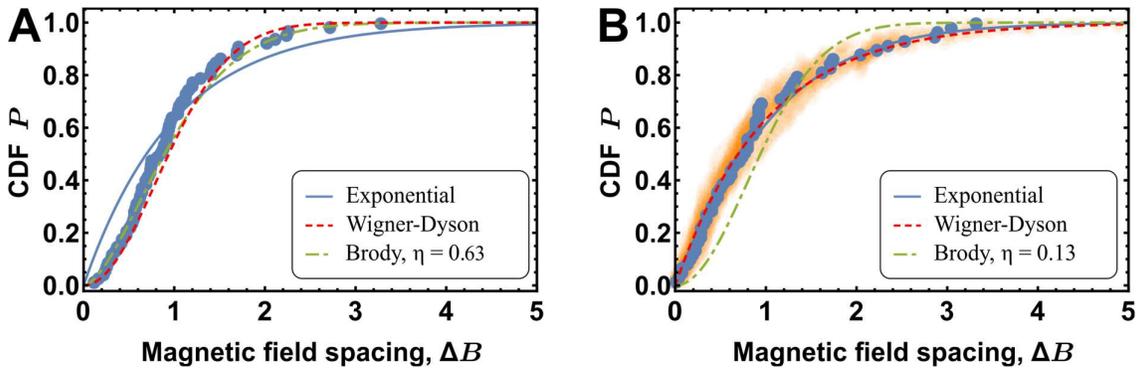

Figure 3. Comparison of NNSFR distributions obtained using RMT simulations for S+D resonances. (A) ECDF of NNSFR for the spectrum measured at 12 μK fitted with Brody distribution ($\eta_{S+D} = 0.63$). (B) ECDF of the spectrum generated by the RMT model for an independent S and D set of resonances with $\epsilon_{S}/h = 5.6$ MHz and $\epsilon_{D}/h = 10.7$ MHz values of mean energy spacing between molecular bound states in corresponding Born-Oppenheimer molecular potentials.

| | Mean bound states energy spacings $\epsilon_{d}$ | Coupling parameter, $v_{cpl}$ | Trapping beam power, W | Magnetic field range | Magnetic field discretization | Number of simulations | Number of bound states | Density, $\rho$ | Brody parameter, $\eta$ |
|---|---|---|---|---|---|---|---|---|---|
| S | 5.6 | 0.7 | 0 | 20 | 2000 | 30 | 500 | 2.1±0.4 | 0.07±0.03 |
| D | 10.7 | 2.9 | 0 | 20 | 2000 | 30 | 500 | 1.0±0.2 | 0.18±0.05 |

Table 1. Summarized RMT simulation parameters and NNSFR analysis results for the case of S and D resonances, alone, without Stark shift(s).

| | Mean bound states energy spacings $\epsilon_d$ | Coupling parameter, $v_{cpl}$ | Trapping beam power, W | Magnetic field range | Magnetic field discretization | Number of simulations | Number of bound states | Density, ρ | Brody parameter, η (RMT) | Brody parameter, η (Experimental) |
|---|---|---|---|---|---|---|---|---|---|---|
| S | 5.6 | 0.7 | 0.15 | 20 | 2000 | 30 | 500 | 2.1±0.4 | 0.1±0.04 | 0.08 |
| S+D | 5.6, 10.7 | 0.7, 2.9 | 0.4 | 20 | 2000 | 30 | 500 | 3.1±0.4 | 0.1±0.03 | 0.63 |

Table 2. Summarized RMT simulation parameters and NNSFR analysis results for the S resonances and the sums of S+D resonances with a Stark shift at 0.15 W and a trapping beam power of 0.4 W, corresponding with experimental conditions and results.

Previous research has proposed that the reason for statistic transformation may be a change in resonances caused by the Stark shift [10]. The tensor polarizability of the thulium atom is quite substantial [21] and, therefore, can naturally repulse resonances. Thus, one could expect some correlations between S and D resonance spectra due to the presence of a Stark shift in the optical dipole trap. To address this possibility, we simulated spectra statistics using trapping beam power values ranging from 0 to 5 W for both S and S+D resonances and used the experimentally determined polarizability of free atoms to calculate shifts in both opened and closed channels. The Brody parameter was fitted for each trapping beam power level. The results (the Brody parameter η versus ODT beam power) are presented in Figure 4. Simulation parameters and results corresponding to experimental conditions are summarized in Table 2.

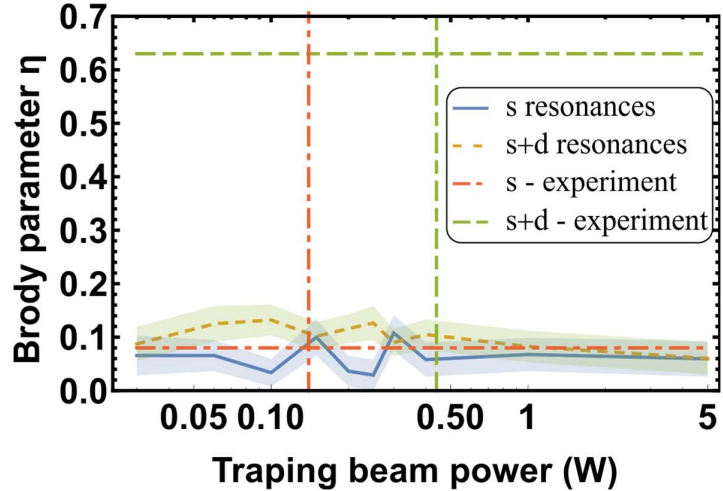

Figure 4 Brody parameters calculated for various trapping beam powers. The blue curve represents s resonances only, and the orange curve represents s and d resonances. The gray area represents the standard deviation of the calculated points. The red and green horizontal lines represent the Brody parameters extracted from the experimental data for S and S+D resonances, respectively [10].

## IV.    RESULTS

Figure 4 shows that, while a noticeable correlation exists with the experimental range of intensities, the Brody parameter for each distribution does not change significantly. Thus, the experimentally observed transformation from random to chaotic behavior could not be explained by modeling of S and D resonances independently. A potential weakness in the above calculation is the use of free atom polarizabilities for both opened and closed channels. It is possible that the molecular polarizability is considerably larger than the atomic polarizability, which would enhance the impact of the Stark effect. However, this scenario is not very likely. In the simulations, the power was varied across a much larger range than that utilized experimentally; so, if the polarizability were underestimated, the effects would be revealed at larger power values. The Brody parameter seems, instead, to decrease, if it changes at all, at high ODT power values (see Figure 4). Therefore, we surmise that there is some other mechanism responsible for the correlation between and/or within the S and D resonance spectra.

## V.    CONCLUSIONS

RMT analyses of the Feshbach spectra of thulium atoms were performed in two cases:  1) changes in resonance density due to the appearance of new resonances with temperature and 2) changes in resonance density due to the Stark shift caused by changes in ODT power. Assuming independent S and D resonances, our calculations were unable to explain the transformations from random to chaotic statistics observed in previously reported experiments. In all the scenarios considered herein, the simulated Brody parameter $\eta \approx 0.1$ is significantly smaller than the experimentally observed value of $\eta = 0.63$. This implies a vital dependence between S and D resonance spectra.

This research was supported by the Russian Science Foundation grant #18-12-00266.